\documentclass[sigconf,nonacm]{acmart} 
\usepackage[english]{babel}
\usepackage{graphicx}
\usepackage{tikz}
\usepackage{pgfplots}
\pgfplotsset{compat=1.18}

\usepackage{listings}
\renewcommand\footnotetextcopyrightpermission[1]{} 
\setcopyright{none}
\AtBeginDocument{%
  }

\begin{document}

\title{Version Control System for Data with MatrixOne}

\author{Hongshen Gou}
\email{gouhongsheng@matrixorigin.cn}
\affiliation{
  \institution{MatrixOrigin}
  \city{Shanghai}
  \country{China}
}

\author{Feng Tian}
\email{tianfeng@matrixorigin.io}
\affiliation{
  \institution{MatrixOrigin}
  \city{San Jose}
  \country{USA}
}

\author{Long Wang}
\email{wanglong@matrixorigin.io}
\affiliation{
  \institution{MatrixOrigin}
  \city{Shenzhen}
  \country{China}
}

\author{Nan Deng}
\email{dengnan@matrixorigin.io}
\affiliation{
  \institution{MatrixOrigin}
  \city{Shanghai}
  \country{China}
}

\author{Peng Xu}
\email{xupeng@matrixorigin.cn}
\affiliation{
  \institution{MatrixOrigin}
  \city{Shanghai}
  \country{China}
}

\begin{abstract}
The rapid advancement of artificial intelligence has elevated data to a cornerstone 
of modern software systems. As data projects become increasingly complex and dynamic, 
version control for data has become essential rather than merely convenient. 

Existing version control systems designed for source code are inadequate for large-scale 
data management, as they often require loading entire datasets into memory for diff and merge 
operations. Database systems, while providing robust data management capabilities, lack 
native support for version control operations such as diff and merge between data forks.

We present a version control system for data implemented in MatrixOne, 
a cloud-native relational database system. Our system leverages MatrixOne's 
immutable storage architecture and multi-version concurrency control (MVCC) to enable 
git-like operations on database tables at scale. The system supports the complete 
spectrum of version control operations—clone, tag/branch, diff, merge, and revert—on 
terabyte-scale datasets with near-instantaneous performance. 

This version control system enables data engineers to adopt established software engineering 
workflows: creating branches for isolated experimentation, submitting pull requests for 
change review, and running CI/CD pipelines, all efficiently and safely. Changes in the 
development environment are isolated from production in both data integrity and computing 
resources. Verified changes can be published to production in atomic transactions, 
ensuring data consistency and avoiding service disruptions.
\end{abstract}

\maketitle

\section{Introduction}\label{sec:intro}

The field of artificial intelligence (AI) is undergoing rapid advancement, driven 
by innovations such as deep learning and large language models (LLMs). 
The development and deployment of these powerful AI systems are fundamentally 
dependent on data.  Each stage of AI development and deployment, from  
initial pre-training and subsequent fine-tuning to in-service prompting, requires 
vast amounts of high-quality data to ensure robust performance and accurate outcomes. 
Consequently, many experts consider data as the fuel that powers the new AI revolution.

The ever-increasing importance of data to AI systems has given rise to the field of 
data engineering.  A substantial amount of development work in an AI project
is actually focused on data processing and feature engineering.  The data used in 
such AI projects has grown to be much more complex and dynamic than ever before.
Engineers need not only to analyze static data, but also to clean and modify data, 
to label data (either by humans or by LLMs), and to manage labels.  An AI project 
is not only about building code, but more and more often becomes a data project.

Version control systems (VCS) are among the most important tools in a software 
engineer's daily work.  
Git \cite{Git} and git services such as GitHub \cite{GitHub} have
become the de facto standard tools for software engineers to manage their code.  
However, for data engineers, there is little help.  Managing large amounts of 
data using a VCS designed for managing code is slow. Tools designed for 
diff and merge of source code often load the entire dataset into memory, which 
is not suitable or even worse, not feasible for large datasets.  
Data engineers often resort to store billions of records in file systems or in 
object stores such as AWS S3 \cite{S3}.  Modifying such large files and collaborating
with a team of engineers on the same dataset is difficult, slow, and error-prone.

Database systems offer much more robust and scalable solutions to managing 
large datasets.  Modifying data using SQL insert, delete, and update statements is 
much easier, faster, and safer than writing bytes to part of a very large file.  
Database transactions can support many users to work on a dataset concurrently.  
Privacy and security are also major concerns in data engineering and database 
systems have mature built-in features like authentication, authorization, role 
based access control (RBAC). 

However, data engineers' workflow with data is very different from conventional
online transaction processing (OLTP) and analytical processing (OLAP) 
workloads of database systems.  Data engineers may prefer to work on an
isolated copy of the dataset, making modifications,
conducting experiments, reverting data to a previous state if experiments fail,
and finally merging the changes back to the original dataset.  Data engineering is 
often a team effort, and team members may need to share partial or incremental 
changes to the dataset.  Changes to datasets need to be reviewed and approved by 
other team members, and finally published and merged into the production environment. 
Publishing changes to the production environment must be fast and transactional
to avoid data corruption and service disruptions in production. 
This entire process is developer-centric, analogous to the well-established 
"branch-and-merge" model used in software engineering where developers 
fork a repository, make modifications in an isolated branch, validate their 
work through a code review processes and continuous integration (CI), 
and subsequently merge the changes back into the main codebase.

MatrixOne \cite{MatrixOne} is a cloud-native relational database system developed 
by MatrixOrigin.  It has a powerful and efficient snapshot system that enables 
version control for large amounts of data.  If we consider a database as a git repository and 
a table as a file in the git repository, MatrixOne can support all day-to-day git operations 
such as clone/branch, push/pull, diff, merge, revert, on terabytes of data almost 
instantly.  MatrixOne allows data engineers to label data, 
to make hypothetical changes to data, to compare and review these changes,
to join or aggregate different versions of data with the full power of SQL,
all without any disruption to existing business applications.  
All these data engineering workloads are totally isolated in both data integrity 
and computing resources from the production environment.  Changes to multiple tables 
can be published to production in one transaction to assure data integrity 
and consistency.

In the rest of this paper, Section \ref{sec:vcop} will first introduce the version control operations
supported by MatrixOne.  We explain the semantics of these operations and 
walk through a typical day-to-day workflow of a data engineer using MatrixOne.
Section \ref{sec:snapshot} will explain the implementation of the snapshot system in 
MatrixOne, followed by the implementation of the version control 
operations like diff/merge in Section \ref{sec:vcopimpl}.  Section \ref{sec:eval} 
will present the experimental evaluation of the version control operations.  
Finally, we will explore some of the future directions and possible improvements.

\section{Related Work}\label{sec:related}
The history of version control systems (VCS) dates back to the earliest practices of 
software engineering \cite{SCCS}\cite{RCS}\cite{CVS}.  Distributed version control 
systems like git \cite{Git} became popular as the complexity of software
projects increased and team sizes grew.  Traditionally, VCS are designed for managing 
source code.  Management of large amounts of data is only an afterthought, usually implemented
by extensions such as git-lfs \cite{GitLFS}.  Such extensions usually only offer storage 
and retrieval of files, but cannot find and resolve conflicts of records in the large file.

Some file systems like ZFS \cite{ZFS} support dedup, snapshot, and restore.
Virtualization systems like VMware can snapshot a virtual disk (VMDK)\cite{VMDK}.
A new VM can run from a clone of VMDK and write changes to the clone.  While a database
system running on top of such storage systems can take advantage of these features,
database systems cannot perform any operations that requires understanding the 
meaning of the changes to the data.

Many database systems support snapshots, recover/restore, and 
PITR (Point In Time Recovery).  However, few systems use these features
to support collaboration of a team of data engineers.  Some database
systems like Snowflake \cite{Snowflake} and Supabase \cite{Supabase} 
support snapshots and clones of databases or tables for development
and testing.  As far as we know, no other database systems except 
MatrixOne offer native support of version control operations like 
diff and merge between forks of a database or table.

\section{Version Control Operations}\label{sec:vcop}
Let us consider table \texttt{T(a int, b varchar, c json)} in a database.  
We can refer to a snapshot of \texttt{T} at timestamp
\texttt{ts} as \texttt{T\{mo\_ts = ts\}}.  For example, users can read a table 
at a specific timestamp using 
\begin{verbatim}
  SELECT * FROM T{mo_ts='2025-09-12 12:34:56'}
\end{verbatim}
MatrixOne supports PITR (Point In Time Recovery) and by default, users can refer
to a snapshot of a table within the most recent 24 hours.  Users can also 
create a named snapshot for a table using the following SQL statement:
\begin{verbatim}
  CREATE SNAPSHOT sn1 FOR TABLE T
\end{verbatim}  
and later refer to the snapshot in SQL using \texttt{T\{snapshot = 'sn1'\}}.  
MatrixOne keeps named snapshot until it is explicitly deleted by users.
In the rest of this paper, we will use a shorter notation 
$T_{sn1}$ for \texttt{T\{snapshot = 'sn1'\}}. 
Viewed from version control's perspective, a timestamp based snapshot
corresponds to a git commit and a named snapshot corresponds to a git tag.

Users can also take a named snapshot of a database, which is the collection of snapshots 
of all tables in the database.  Snapshot, clone, and restore operations can be performed
on a database as well.  Later in this paper we will only discuss snapshots of tables.

Users can create a new table by cloning a table from a snapshot \texttt{sn1} using 
\begin{verbatim}
  CREATE TABLE TClone FROM SNAPSHOT T{snapshot='sn1'}
\end{verbatim}
The result table \texttt{TClone} has the same schema (including 
the primary key definition) and data.  Cloning a table roughly 
corresponds to cloning a git repository or creating a new branch 
from a git tag from version control's perspective.  
Once cloned, \texttt{T} and \texttt{TClone} are 
separate tables and users can perform insert, delete, and update 
operations on \texttt{T} and \texttt{TClone} 
independently.  

Now let us look at an example workflow of a data engineer using MatrixOne
for data version control operations.   After creating \texttt{TClone}, 
users can modify data in both \texttt{T} and \texttt{TClone}, 
data in both \texttt{T} and \texttt{TClone}, and create new 
snapshot \texttt{sn2} on \texttt{T} and snapshot \texttt{sn3} 
on \texttt{TClone}.  Later, users want to merge changes 
from $TClone_{sn3}$ back to \texttt{T} to get 
\texttt{sn4}.  This workflow is shown in Listing \ref{lst:wf}.
\begin{lstlisting}[label=lst:wf,caption=Branching and Merging Workflow]
  T:       --> sn1 --> sn2 ------>  sn4 --> 
                \             /
  TClone:        \---> sn3 --/

         ------------- now  ---------> time
\end{lstlisting}
In this example workflow, we say that $T_{sn2}$ and $TClone_{sn3}$ 
share a common base revision $T_{sn1}$. 

Like git, users can push/pull changes from one table to another by 
restoring a snapshot.  For example, users can pull the more recent 
changes in $T_{sn2}$ to table \texttt{TClone} using 
\begin{verbatim}
  RESTORE TABLE TClone FROM
        SNAPSHOT T{snapshot='sn2'}
\end{verbatim}
Restore will overwrite all modifications in $TClone_{sn3}$ and completely replace 
data of \texttt{TClone} with data of $T_{sn2}$.  It is equivalent to perform 
\texttt{git reset --hard sn2} on \texttt{TClone}.

Users can diff two snapshots using 
\begin{verbatim}
  SNAPSHOT DIFF T{snapshot='sn2'} 
         AND TClone{snapshot='sn3'}
\end{verbatim}  
\texttt{SNAPSHOT DIFF} does not require that the two snapshots 
were branched from a common base revision, as long as they have the same schema. 
That is, the two snapshots must have the same column names and column types in the 
same order and same primary key definitions if they have one.

\texttt{SNAPSHOT DIFF} treats tables as an unordered multi-sets of records.
Logically, its result is the same as the result of the following SQL query 
in Listing \ref{lst:diff}.  Later in the paper we will see that when  
two snapshots share a commmon base revision, MatrixOne can perform 
diff and merge between them very efficiently.
\begin{lstlisting}[label=lst:diff,caption=SNAPSHOT DIFF Query]
  WITH UnionT as (
    SELECT -1 as cnt, a, b, c 
    FROM T{snapshot='sn2'}
    UNION ALL
    SELECT 1 as cnt, a, b, c 
    FROM TClone{snapshot='sn3'}
  )
  SELECT sum(cnt) as diffCnt, a, b, c FROM UnionT 
  GROUP BY a, b, c
  HAVING sum(cnt) <> 0
\end{lstlisting}
Each row in the result of \texttt{SNAPSHOT DIFF} 
represents a potential conflict between the two snapshots.

Users can perform a three-way merge from source table 
\texttt{TClone} (currently at snapshot \texttt{sn3}) 
into target table \texttt{T} (currently at snapshot \texttt{sn2}) 
using $T_{sn1}$ as the common base revision. 
\begin{lstlisting}[label=lst:merge,caption=SNAPSHOT MERGE Query]
  SNAPSHOT MERGE TABLE T FROM TClone{snapshot='sn3'}
    [BASED ON T{snapshot='sn1'}]
    [WHEN CONFLICT FAIL|SKIP|ACCEPT]
\end{lstlisting}
We allow users to optionally specify a snapshot of \texttt{TClone} for merge source,
but the merge target must be the current version of a table.
The \texttt{BASED ON} clause is optional and if not specified, MatrixOne will try to 
find an implicit common base revision between merge source and target.  
Merge is different from restoring a snapshot in 
that merge will resolve conflicts found in the result of \texttt{SNAPSHOT DIFF}.
MatrixOne supports three modes of conflict resolution: \texttt{FAIL}, \texttt{SKIP}, 
and \texttt{ACCEPT}.  \texttt{FAIL} means that merge will fail if conflicts are found
and users must resolve all the conflicts manually before merging can proceed.  
\texttt{SKIP} means that merge will resolve the conflicts by keeping 
the version in merge target \texttt{T}, and \texttt{ACCEPT} means that merge 
will accept the version in merge source \texttt{TClone}.  

Conflict resolution between snapshot $T_{sn2}$ and 
snapshot $TClone_{sn3}$ is computed differently depending on whether tables 
\texttt{T} and \texttt{TClone} have primary keys. The presence of primary keys 
enables more efficient conflict detection by uniquely identifying rows, whereas 
tables without primary keys require matching on all column values.

\paragraph{Conflict Resolution with Primary Keys.}
Consider the case where \texttt{T} and \texttt{TClone} have 
the same primary key on column \texttt{a}. When \texttt{SNAPSHOT DIFF}
identifies a potential conflict on a row with primary key \texttt{a\_value}, 
we must determine whether this represents a true conflict (both branches modified 
the same row differently) or a false conflict (only one branch modified the row, 
or both branches made compatible changes). 

To make this determination, we examine the presence of \texttt{a\_value} in three 
snapshots: the common base revision $T_{sn1}$, the target snapshot $T_{sn2}$, 
and the source snapshot $TClone_{sn3}$. The following six scenarios cover all 
possible cases:

\begin{enumerate}
\item \texttt{a\_value} does not exist in the common base revision $T_{sn1}$ 
and \texttt{a\_value} does not exist in $TClone_{sn3}$, but exists in $T_{sn2}$.
This means that a row with primary key \texttt{a\_value} is inserted into
$T_{sn2}$.  This is a false conflict and merge will keep the row in $T_{sn2}$.

\item \texttt{a\_value} does not exist in $T_{sn1}$
and does not exist in $T_{sn2}$, but exists in $TClone_{sn3}$.
This indicates that only $TClone_{sn3}$ inserted the row with primary key 
\texttt{a\_value} after branching. This is a false conflict and 
merge retains the row from $TClone_{sn3}$.

\item \texttt{a\_value} does not exist in the common base revision 
$T_{sn1}$ and \texttt{a\_value} exists in both $T_{sn2}$ and $TClone_{sn3}$.
This means both $T_{sn2}$ and $TClone_{sn3}$ have inserted the row 
with primary key \texttt{a\_value} but with different values in 
some other columns.  This is a true conflict.  
\texttt{SKIP} mode will use the version in $T_{sn2}$, 
\texttt{ACCEPT} mode will use the version in $TClone_{sn3}$,
and \texttt{FAIL} modewill cause the merge to fail.

\item \texttt{a\_value} exists in the common base revision $T_{sn1}$ 
and in $TClone_{sn3}$, and the values of all columns of the two rows
are the same.  This means this row is not changed in $T_{sn2}$ and 
$TClone_{sn3}$ has deleted or updated this row.  This is a false 
conflict and merge will perform the same operation as performed 
in $TClone_{sn3}$.

\item \texttt{a\_value} exists in the common base revision $T_{sn1}$ 
and in $TClone_{sn3}$, and the values of all columns of the row 
are the same. This means this row is not changed in $TClone_{sn3}$
and only $T_{sn2}$ has deleted or updated this row. 
This is a false conflict and merge will perform the 
same operation as performed in $T_{sn2}$.

\item \texttt{a\_value} exists in $T_{sn1}$,
but the row values in both $T_{sn2}$ and $TClone_{sn3}$ differ from $T_{sn1}$ 
(and potentially one of them is a deletion). This indicates that both branches 
independently modified the same row differently. This is a true conflict. 
\texttt{SKIP} retains the version 
from $T_{sn2}$ (which may be a deletion), \texttt{ACCEPT} retains the version from 
$TClone_{sn3}$ (which may also be a deletion), and \texttt{FAIL} aborts the merge.
\end{enumerate}

The key insight is that a true conflict occurs only when both branches independently 
modify the same row. Cases where only one branch modifies a row (scenarios 1, 2, 4, and 5) 
are false conflicts that can be automatically resolved by applying the modification from 
the appropriate branch.

\paragraph{Conflict Resolution without Primary Keys.}
When tables \texttt{T} and \texttt{TClone} do not have a primary key,
rows cannot be uniquely identified. For each set of rows with identical values 
in all columns identified by \texttt{SNAPSHOT DIFF}, we must consider the 
cardinality (count) of such rows in each snapshot. 

Let $N_{sn1}$, $N_{sn2}$, and $N_{sn3}$ denote the number of rows with identical 
values in all columns in the common base revision $T_{sn1}$, snapshot $T_{sn2}$, 
and snapshot $TClone_{sn3}$, respectively. Let $\delta_T = N_{sn2} - N_{sn1}$ 
represent the net change in the target branch (positive for inserts, negative for 
deletes), and $\delta_{TClone} = N_{sn3} - N_{sn1}$ represent the net change in 
the source branch. The following cases must be considered:
\begin{enumerate}
\item $\delta_T = 0$ indicates that the target branch 
made no net change to rows with these values. This is a false conflict and 
merge retains $N_{sn3}$ such rows (applying the source branch's changes).

\item $\delta_{TClone} = 0$ indicates that the source 
branch made no net change. This is a false conflict and merge retains $N_{sn2}$ 
such rows (applying the target branch's changes).

\item $\delta_T \neq 0$ and $\delta_{TClone} \neq 0$ 
indicates that both branches independently modified the count of rows with these 
values. This is a true conflict. \texttt{SKIP} mode retains $N_{sn2}$ such rows, 
\texttt{ACCEPT} mode retains $N_{sn3}$ such rows, and \texttt{FAIL} mode aborts the merge.
\end{enumerate}

In case 3, alternative resolution strategies are possible, such as retaining 
$max(N_{sn2}, N_{sn3})$ rows. However, MatrixOne treats this as a true conflict 
requiring explicit resolution to prevent accidental data loss.

\section{Snapshots in MatrixOne}\label{sec:snapshot}
MatrixOne implements data version control operations using its snapshot system.
In this section, we will first briefly explain the architecture of MatrixOne
database and its storage and transaction management system,
then explain how the snapshot system works.

\begin{figure}[htbp]
  \centering
  \includegraphics[width=\linewidth]{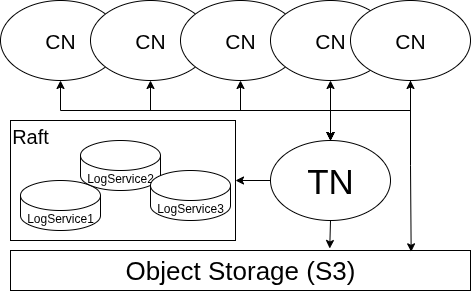}
  \caption{The Architecture of MatrixOne Database}
  \Description{MatrixOne Database Architecture}
  \label{fig:m1a}
\end{figure}

MatrixOne is a cloud-native, distributed database system that supports 
both transactional and analytical workloads (HTAP workload).  A MatrixOne database 
system has three kinds of nodes as shown in Figure \ref{fig:m1a}.
\begin{description}
\item[LogService] nodes form a Raft \cite{Raft} group and is responsible for 
storing the write ahead log (WAL) of the database system.  A MatrixOne database 
system usually has 3 or 5 LogService nodes.
\item[TN] node is the transaction decision node which decides if a transaction can commit.  TN 
node serializes committed transaction logs.  TN node also acts as a hub of a pub/sub system 
of WAL, streaming WAL records of tables to CNs that have subscribed to the tables.  
A MatrixOne database system usually has only one TN node because TN node 
is stateless and can be recovered from LogService nodes in just a few seconds.  
However, if necessary, the TN node can have a hot standby node. 
\item[CN] nodes are the compute nodes that execute SQL queries.  A MatrixOne 
system can have an unlimited number of CN nodes.  A single SQL query may be 
executed on multiple CN nodes in parallel, but simple queries like insert or 
update a few rows are usually executed on one CN node.
\end{description}

MatrixOne stores data of a table in an object storage system such as S3 \cite{S3}.  
Each object contains a few row groups and each row group stores many rows 
using column store format \cite{ColumnStore}.  
Once written to object storage, the object is immutable.  
Objects of a table form an LSM tree \cite{LSMTree}, 
ordered by the primary key columns or, if without primary key, by clustering key columns
with a uniquifier.  Each row also has a physical rowid that consists of the name of the 
object and the position of the row in the object.
To delete a row, a tombstone record that contains the primary key 
(or clustering key plus uniquifier) and the physical rowid
of the deleted row is written to a separate object.
The metadata of all the objects of a table is stored in a directory structure.

MatrixOne implements a MVCC system similar to that of the PostgreSQL database \cite{PostgreSQL}.  
All transactions begin execution in a CN node.  When a transaction needs to 
read or write a table, CN registers a subscription to the table to TN and
receives up-to-date WAL records of the table.  All transactions in one CN node
share the same subscription to one table and the subscription may be kept
alive for a while after the transaction commits for immediate reuse by 
another transaction.
Each transaction sees a consistent snapshot of the database and stores 
all modifications in its own transaction workspace.  
When the data in the workspace grows beyond a threshold, CN 
writes the data to the object storage bypassing TN and only stores 
the metadata of the written objects in the workspace.  
When it is time to commit the transaction, CN writes the workspace to TN and 
TN commits the transaction by writing WAL to LogService.  
Committed WAL records are streamed to all CN nodes that have 
subscribed to the table.  Workspace and commit WAL records may consist of
a mix of changed rows and metadata of objects, therefore are memory efficient 
for large transactions like loading data or modifying a large number of rows.

If a transaction only modifies a few rows, TN will commit that transaction by
writing WAL records to LogService node, then will write these rows  
to an in-memory object without writing to object storage.  This in-memory object
can accumulate small writes from many transactions.  Rows in the in-memory 
object have transaction timestamps and TN can only append to, but not modify 
the in-memory object.  This in-memory object logically is still immutable up 
to the most recently committed transaction.  When enough rows are accumulated 
or a certain time has passed, TN will close this in-memory object and write 
it to object storage.  

Each CN has a local cache of S3 objects.  The cache implementation 
is extremely simple due to the immutable nature of objects.  Each CN
also rebuilds a copy of the in-memory object of TN by reading
and applying the WAL records from the subscription of the table.

A snapshot of a table is simply the directory structure of 
the metadata of all objects (both in memory objects and 
on storage objects) of the table.  Reading of a timestamp-based 
snapshot is implemented by reading objects in the directory 
structure and applying proper timestamp filter on records as required 
by the MVCC system.  To save a named snapshot, MatrixOne forces 
a flush of the in-memory objects of the table to object storage.  
The system stores the directory structure of metadata of objects 
and does necessary bookkeeping to associate the name of the 
snapshot with the directory structure.

Like other typical implementations of log-structured storage 
systems \cite{LSMTree}\cite{LSFS}, MatrixOne uses
background compaction and garbage collection processes
to compact objects and reclaim storage space.  The garbage 
collection process understands the snapshot system and will 
not delete objects that are referenced by a named snapshot.

\section{The Implementation of Version Control Operations}\label{sec:vcopimpl}
With the snapshot system in MatrixOne, cloning and restoring a table 
from a snapshot is relatively easy. 
Cloning a table from a snapshot is implemented by copying the 
directory structure of metadata of objects of the snapshot. 
Restoring a table from a snapshot is simply setting 
the current state of the table to the snapshot.

Next, we describe the implementation of the diff and merge operations.  
Consider the workflow on table \texttt{T} and \texttt{TClone} in Listing 
\ref{lst:wf}.   After creating \texttt{TClone}, users modified 
data in both \texttt{T} and \texttt{TClone} and the two tables 
have progressed independently to different snapshots 
\texttt{sn2} and \texttt{sn3}.  We use $\Delta_{sn2}$ to denote 
the set difference of objects in snapshot $T_{sn2}$ and in the 
common base revision $T_{sn1}$.  $\Delta_{sn2}$ contains objects 
that are added to $T_{sn2}$ due to data modifications. Especially, 
data deletions are implemented as additions of tombstone records. 
$\Delta_{sn2}$ may also contain objects only in $T_{sn1}$ due to 
compaction and garbage collection.  
Similarly, we use $\Delta_{sn3}$ to denote the set difference 
of objects in $TClone_{sn3}$ and $T_{sn1}$.

\subsection{Diff}\label{sec:diff}
To find \texttt{SNAPSHOT DIFF} of $T_{sn2}$ and $TClone_{sn3}$, 
we only need to read the objects in $\Delta_{sn2}$ and $\Delta_{sn3}$.  
If table \texttt{T} has a primary key \texttt{a},
all operations on one primary key in $\Delta_{sn2}$ can be collapsed 
into one logical operation: a delete from $T_{sn1}$, an insert to $T_{sn2}$, 
or an update (in this case, the logical operation consists 
of two physical operations, a delete followed by an insert).
Note that the physical operation of deletion is always performed 
on a row in the common base revision $T_{sn1}$.  
This scanning and collapsing read operation on $\Delta_{sn2}$ is the same 
as an ordinary table scan on the LSM tree and applying tombstone records,
except that instead of masking off deleted rows,  we need to scan out the 
deletion operation of a row as well.  
We give a minus sign to the deletion operation and a plus sign to 
the insert operation in the result of scanning $\Delta_{sn2}$.

We want to point out some differences between scanning $\Delta_{sn2}$
and the SQL query in Listing \ref{lst:diff}.  
First, the plus and minus sign in scanning $\Delta_{sn2}$ 
is used to indicating difference between $T_{sn2}$ and $T_{sn1}$, 
while the plus and minus sign in Listing \ref{lst:diff} is used to 
indicate the difference between $T_{sn2}$ and $TClone_{sn3}$.
Second, for deleted rows, $\Delta_{sn2}$ only scans out the 
tombstone records.  The non-primary key columns of the tombstone records 
are filled with nulls.  Later, we will join with $T_{sn1}$ to retrieve
the real values of the deleted row, only if needed.

$\Delta_{sn3}$ is scanned in the same way.

Next we will perform a special diff aggregation to find the differences
of $\Delta_{sn2}$ and $\Delta_{sn3}$.  Changes in $\Delta_{sn2}$ 
and $\Delta_{sn3}$ are considered the same if they are deletions 
on the same row in $T_{sn1}$, or, insertions of rows with the same 
values in all columns.  These same changes on both sides are 
cancelled out by the aggregation.  With the result of the diff aggregation,
\texttt{SNAPSHOT DIFF} will compute the result by,
\begin{enumerate}
\item flipping the sign of rows from $\Delta_{sn2}$. 
\item for the tombstone records in the aggregation result, we perform 
a lookup (a join with $T_{sn1}$) to find out the values of non primary 
key columns of the row.
\end{enumerate}

If table \texttt{T} has no primary key, we perform the same scan on 
$\Delta_{sn2}$ and $\Delta_{sn3}$ and the same diff aggregation as described 
above.  The diff aggregation cancels out inserted rows to $T_{sn2}$ and 
$TClone_{sn3}$ with the same values in all columns and deletions 
with the same physical rowid (the tombstone record contains uniquifier and physical rowid).  
Then we lookup values of all columns of deleted rows using physical rowid.
The result of \texttt{SNAPSHOT DIFF} is computed using the query in 
Listing \ref{lst:diff} on the diff aggregation result of $\Delta_{sn2}$ and $\Delta_{sn3}$.

\subsection{Three Way Merge} \label{sec:threewaymerge}
A three-way merge from \texttt{TClone} to \texttt{T} is implemented by first 
performing the scan and diff aggregation on $\Delta_{sn2}$ and $\Delta_{sn3}$, 
as described in Section \ref{sec:diff}. The plus and minus signs in the diff 
aggregation indicate whether a row was inserted into a snapshot or deleted from the
common base revision. 

\paragraph{Merge with Primary Keys.}
When table \texttt{T} has a primary key, we use the diff aggregation result to 
determine conflict types. A potential conflict is a \textbf{true conflict} if the 
primary key appears in both $\Delta_{sn2}$ and $\Delta_{sn3}$ (indicating both 
branches modified the same row). It is a \textbf{false conflict} if the primary key 
appears in only one of $\Delta_{sn2}$ or $\Delta_{sn3}$ (indicating only one branch 
modified the row).

\paragraph{Handling Row Movement.}
A special case occurs when one branch updates a row in the common base revision to 
have identical values in all columns but at a different position (e.g., due to 
LSM tree reorganization). This is effectively a "move" operation rather than a 
data change. We treat conflicts caused by such moves as false conflicts to avoid 
incorrectly rejecting valid updates from the other branch.

For example, if a row is moved in $T_{sn2}$ (same values, different position) and 
updated in $T_{sn3}$, without special handling, the \texttt{SKIP} strategy would 
treat this as a true conflict and keep the moved row from $T_{sn2}$, losing the 
update from $T_{sn3}$. By detecting that the row values are unchanged in $T_{sn2}$ 
(only position changed), we correctly identify this as a false conflict and apply 
the update from $T_{sn3}$.

The move case corresponds to scenario 6 in Section \ref{sec:vcop} and is the only 
case where merge needs to read the full deleted row from the common base revision 
to compare values. In practice, such moves are rare (typically occurring only after 
compaction), so this lookup is seldom needed.

\paragraph{Merge without Primary Keys.}
When table \texttt{T} does not have a primary key, we perform the same scan and 
diff aggregation. A potential conflict is a true conflict only if a row with identical 
values in all columns appears in both $\Delta_{sn2}$ and $\Delta_{sn3}$ (indicating 
both branches modified rows with these values). In this case, merge reads all columns 
of deleted rows from the common base revision using physical rowid to determine the 
conflict resolution strategy as described in Section \ref{sec:vcop}.

\subsection{Two Way Merge}
Users need not specify the common base revision in the optional 
\texttt{BASED ON} clause of the \texttt{SNAPSHOT MERGE} statement.  
In most cases, MatrixOne maintains proper bookkeeping of snapshot and clone 
lineages, enabling the system to automatically identify the common base revision 
of \texttt{T} and \texttt{TClone}. A two-way merge is implemented as a three-way 
merge with an implicit common base revision.

Sometimes the system may not be able to know the common base revision or 
the common base revision is not available.  For example, two tables are clones 
from the same original table and then users deleted the original table and 
all its snapshots.  In this case, the two-way merge is computed as a three-way
merge with an empty common base revision.  Even in this case, if the two tables
are in fact cloned from a common ancestor, they will likely share many common 
objects.  The diff aggregation can be computed much more efficiently than using 
the SQL query in Listing \ref{lst:diff} of Section \ref{sec:vcop} by simply 
skipping the objects common to both $T_{sn2}$ and $TClone_{sn3}$.

\subsection{Compaction and Garbage Collection}
MatrixOne does not compact or garbage collect objects that 
are referenced by named snapshots. However, the system may
schedule compaction or garbage collection (GC) jobs on table \texttt{T} between 
\texttt{sn1} and \texttt{sn2}, or on \texttt{TClone} between \texttt{sn1} and \texttt{sn3}.  

MatrixOne performs compaction or GC as a transaction that deletes old objects and 
writes new objects, effectively moving all valid rows from several old objects to 
one or more new objects. This reorganization can cause rows to appear at different 
positions even when their values are unchanged.

Users typically branch from well-organized snapshots, so compaction of objects
in the common base revision is rare. When compaction does occur, the merge algorithm 
handles it correctly: the diff aggregation described in Section \ref{sec:threewaymerge} 
identifies rows that were only moved (same values, different positions) and treats 
them as unchanged, preventing false conflicts.

\subsection{Discussion} \label{sec:discussion}
We discuss some interesting issues and possible future works related to 
the implementation of version control operations of MatrixOne.

\subsubsection{Three Way Diff}
Currently, MatrixOne only supports two-way diff with a common base revision.  
A general three-way diff can be implemented efficiently by allowing users to specify 
a \texttt{BASED ON} clause in the \texttt{SNAPSHOT DIFF} statement. In fact, the plus 
and minus signs in the diff aggregation contain exactly the information needed to compute 
three-way diff. While a general three-way diff is useful for understanding potential 
conflicts in complex histories, we believe end users performing tasks such as data change 
review will most likely work with two-way diff results. Therefore, we decided not to expose 
three-way diff to reduce operational complexity and potential user confusion.

\subsubsection{Revision Lineage}
MatrixOne keeps track of the revision lineage of two tables that are one 
branch away from each other.  When a two-way merge cannot find a common 
base revision, an empty common base revision is used.  MatrixOne can 
still optimize the merge by skipping shared objects.  This is different 
from a three-way merge that users explicitly specify the common base revision.
Without a common base revision, MatrixOne cannot tell whether a row is newly 
inserted or updated -- especially, it cannot tell if the row is simply "moved".
To be safe, MatrixOne can only treat conflicts caused by "move" as true 
conflicts and ask 
users to decide how to resolve the conflicts.

\subsubsection{Conflict Resolution}
MatrixOne only supports SKIP (accept yours in git merge) and ACCEPT (accept mine)
modes for conflict resolution.  If users do not want to use these modes, users
must resolve the conflicts manually.  Users must find out all conflicts using
\texttt{SNAPSHOT DIFF} and then modify data in their own table \texttt{TClone} 
using SQL.  MatrixOne considers conflicts at row level, not at cell level.  
When both $\Delta_{sn2}$ and $\Delta_{sn3}$ modify the same row, it is a 
conflict even if the modifications are on different, non-primary key columns.  
We may consider relax this rule to allow users to automatically resolve 
conflicts at cell level in the future.

\subsubsection{Indices}
Cloning a table creates a new table with the same schema and primary key definitions.
Currently, cloning a table does not clone secondary indices.  
Secondary indices in MatrixOne are implemented using auxiliary tables
consisting of the indexed columns and the primary key columns of 
the original table. The secondary index is another table stored and managed as
an LSM tree, similar to user tables in MatrixOne. In the future, we may consider cloning 
secondary indices by cloning the auxiliary index tables.  

\subsubsection{Large Object Types}
MatrixOne supports two kinds of large object types, LOB and datalink.
The first kind includes TEXT, JSON, BLOB types and these types are 
stored inside the database.  There is no difference between these 
LOB types and other data types when we perform diff or merge operations,
except that they may consume a lot of memory if we hold the full contents
of these LOBs in the hash table of the join or diff aggregate operators. 
MatrixOne solves this memory consumption problem by building the hash 
table on a signature such as SHA256 of the LOB and releasing the memory 
of the full contents of the LOB.  The second kind of large object types 
is the datalink type.  A datalink is basically a URL pointing to an 
external resource, for example, a file in a network file system or 
an object in an object storage system like S3.  MatrixOne does not 
manage changes of the content of the external resource.  A datalink value 
is changed only if the URL is changed.

\subsubsection{Schema Change}
Users can make schema changes on a table using \texttt{ALTER TABLE} 
statement.  In particular, MatrixOne supports \texttt{RESTORE TABLE} 
to a snapshot that was taken before the schema change.  However, 
if users alter the schema of a table or a cloned table, MatrixOne
will not be able to perform diff or merge between the two tables.  
To use data version control on such a table, it is generally 
advised to make schema changes on a table before cloning it. 

\section{Experimental Evaluation}\label{sec:eval}
The experiments are run on a Linux (kernel \texttt{5.19.12}) server
with Intel Xeon Silver CPU (2.4 GHz, 64 cores), 256~GB memory, 
and local SSD disks.  The MatrixOne database system is installed 
on a Kubernetes cluster.  We evaluate MatrixOne's data version 
control operations using the lineitem table of the TPC-H 
100GB dataset \cite{TPCH}.  
The lineitem table is the largest table in the dataset and contains 
about 600 million rows.  We will conduct experiments with 
two scenarios, one with \texttt{(l\_orderkey, l\_linenumber)} as 
the primary key (later referred to as the PK test cases), and one without 
a primary key (NoPK test cases).  Even without defining a primary key,
data into the lineitem table is still clustered in the order of
\texttt{(l\_orderkey, l\_linenumber)} after loading the table.

\subsection{Experiment 1: Clone}\label{sec:expClone}
Users can create a new copy of the data by cloning the lineitem table,
or by creating a new table T and inserting the data into it
using the following SQL query,
\begin{verbatim}
  INSERT INTO T SELECT * FROM lineitem;
\end{verbatim}
Table \ref{tab:clone} shows the cost of clone vs insert, both in terms of time 
and storage space.  Cloning from a snapshot does not require 
reading the full data of the table and only need 
to create a copy of the metadata directory structure.  
Inserting using SQL will read all the data of the lineitem table and write 
to new table T, incur 34 GB of additional storage space.  
Primary key of a table is implemented as the sorting order 
of the LSM tree in MaxtrixOne, therefore does not incur additional 
storage overhead.  MatrixOne also carefully optimized batch insert of 
large amount of data into a table with primary key.  For the case that
the target table is empty, checking primary key constaint is a largely free.

\begin{table}[h]
  \centering
  \caption{Clone vs Insert}
  \label{tab:clone}
  \begin{tabular}{lcc}
    \toprule
    \textbf{Operation} & \textbf{Time (s)} & \textbf{Space} \\
    \midrule
    Clone, PK  & 0.20 & 314 KB \\
    Clone, noPK & 0.17 & 294 KB \\
    Insert, PK  & 114.6 & 34 GB \\
    Insert, noPK & 119.3 & 34 GB \\
    \bottomrule
  \end{tabular}
\end{table}

\subsection{Experiment 2: Diff and Merge}
Once cloned the lineitem table to table T, we modified data in 
table T using SQL.  We conducted the experiment using the following
four change sets of different sizes,
\begin{itemize}
\item \textbf{C1}: update 1{,}000 random rows.
\item \textbf{C2}: update 10{,}000 random rows.
\item \textbf{C3}: update 100{,}000 random rows.
\item \textbf{C4}: update 1{,}000{,}000 random rows.
\end{itemize}
MatrixOne maintains the sorting order of LSM tree when \texttt{T} has a primary key.
Comparing to the case that \texttt{T} does not have a primary key,
searching a few rows using primary key is usually much faster, but updating many rows
can generate smaller, more fragmented objects.  The next round of compaction and GC 
will reduce the fragmentation.

After applying these changes, we compute the diff between T and the original 
lineitem table, and merge the changes from T to the original lineitem table
using \texttt{ACCEPT} mode.  We compare our built-in snapshot diff/merge 
operations with an SQL implementation.  The SQL query for diff is in 
Listing \ref{lst:diff} of Section \ref{sec:diff} and the SQL query for 
merge is in Listing \ref{lst:merge-sql}.

\begin{lstlisting}[label=lst:merge-sql,caption=SQL for Merge]
CREATE TABLE Diff AS
    ... the diff SQL query of ...
    ;

DELETE FROM lineitem
WHERE (l_orderkey, l_linenumber) IN (
  SELECT l_orderkey, l_linenumber FROM Diff 
  WHERE diffCnt < 0
);

INSERT INTO lineitem 
SELECT l_orderkey, l_linenumber, ... 
FROM Diff WHERE diffCnt > 0
;
\end{lstlisting}

We want to point out that the SQL implementations in the case of without 
a primary key actually cheated a bit by assuming the external 
knowledge that \texttt{(l\_orderkey, l\_linenumber)} is in fact
the primary key.  This is commonly practiced in OLAP systems to avoid 
the overhead of checking and maintaining primary key constraints.  
Without this assumption, we need to write a SQL query to delete 
only a certain number of rows from all duplicate rows, this query 
is not trivial and will be much slower to execute.

\begin{table}[h]
  \centering
  \caption{Diff, BuiltIn vs.\ SQL Time(s)}
  \label{tab:expdiff}
  \begin{tabular}{lcccc}
    \toprule
    \textbf{Operation} & \textbf{C1} & \textbf{C2} & \textbf{C3} & \textbf{C4} \\
    \midrule
    BuiltIn Diff, PK  & 0.19 & 0.38 & 1.73 & 3.27 \\
    BuiltIn Diff, NoPK  & 0.85 & 22.50 & 9.04 & 60.19 \\
    SQL Diff, PK  & 316.16 & 418.19 & 428.78 & 431.50 \\
    SQL Diff, NoPK  & 378.19 & 396.61 & 394.13 & 371.74 \\
    \bottomrule
  \end{tabular}
\end{table}

Table \ref{tab:expdiff} shows the execution time of built-in diff and SQL diff operations.
Built-in diff operations are significantly faster than SQL implementations in all cases, 
achieving 100-500x speedup. The performance advantage stems from scanning only the 
$\Delta$ objects (changed data) rather than entire tables. Built-in diff operations 
are implemented as optimized queries that leverage the query optimizer. 

The performance difference between C2 and C3 in the NoPK case reflects the optimizer 
selecting different parallelization strategies when crossing a threshold in the 
amount of data being processed. The non-monotonic behavior (C2 taking longer than C3) 
is due to the optimizer choosing a less efficient strategy for the intermediate size.

\begin{table}[h]
  \centering
  \caption{Merge, BuiltIn vs.\ SQL Time(s)}
  \label{tab:expmerge}
  \begin{tabular}{lcccc}
    \toprule
    \textbf{Operation} & \textbf{C1} & \textbf{C2} & \textbf{C3} & \textbf{C4} \\
    \midrule
    BuiltIn Merge, PK  & 0.35 & 0.97 & 7.95 & 16.13 \\
    BuiltIn Merge, NoPK  & 0.88 & 22.70 & 12.09 & 68.75 \\
    SQL Merge, PK  & 321.52 & 412.89 & 442.68 & 471.16 \\
    SQL Merge, NoPK  & 393.95 & 405.35 & 401.59 & 403.18 \\
    \bottomrule
  \end{tabular}
\end{table}

Generating the diff set accounts for the majority of merge execution time.
As expected, Table \ref{tab:expmerge} shows that built-in merge operations are 
significantly faster than SQL implementations in all cases, achieving similar 
speedups as diff operations.

From the SQL diff and merge performance, we observe that primary keys improve 
performance with smaller change sizes but introduce overhead as change sizes grow, 
due to index maintenance costs. However, built-in diff and merge operations 
consistently perform better with primary keys than without, for two reasons:
\begin{itemize}
\item \textbf{More efficient lookups:} Primary keys enable direct row identification 
and faster conflict detection.
\item \textbf{Operation collapsing:} The scanning and diff aggregation on $\Delta$s 
can collapse multiple operations on the same primary key into a single logical operation, 
skipping unnecessary checks. For example, an inserted row from scanning $\Delta$ either 
has a matching deleted row (indicating an update) or is a valid insert without conflict.
\end{itemize}

\subsection{Experiment 3: Collaborative Workload Without Conflicts}\label{sec:expcollab}
\begin{figure}[h]
  \centering
  \includegraphics[width=0.9\linewidth]{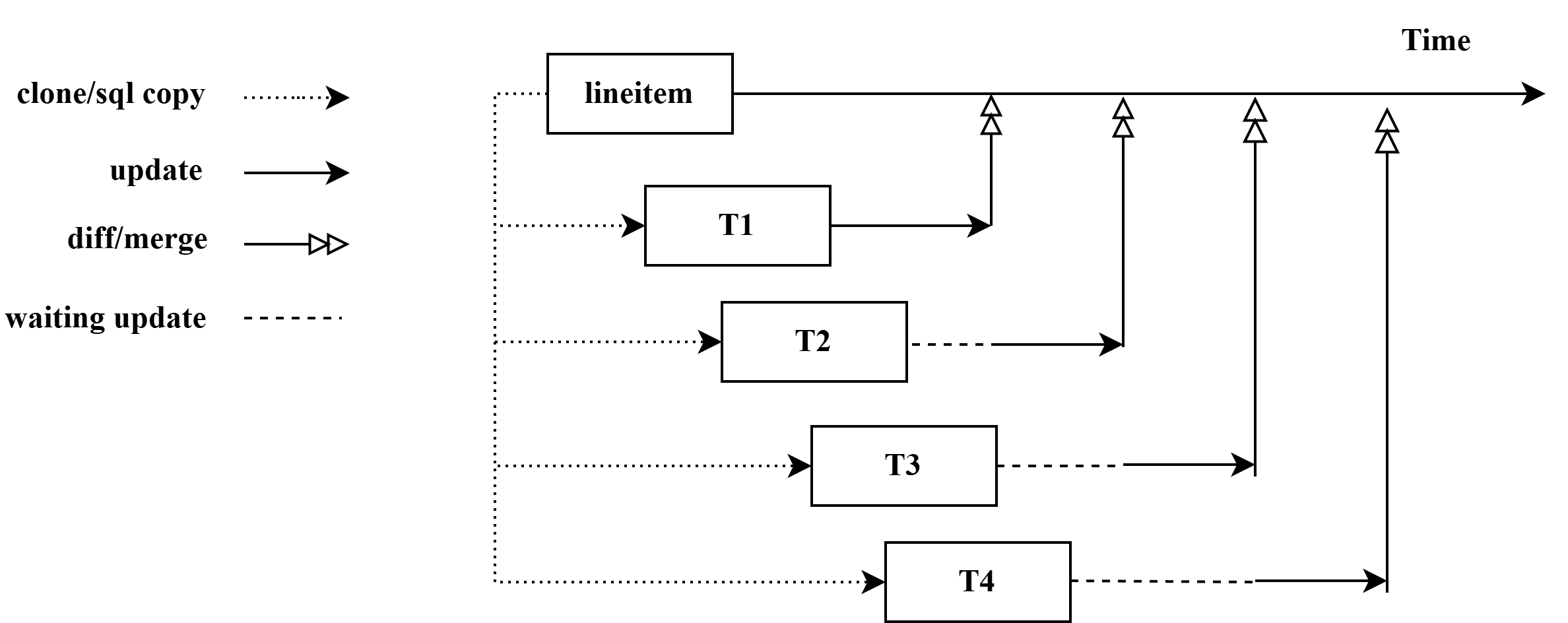}
  \caption{Concurrent Snapshot Diff/Merge Timeline}
  \Description{Timeline showing four concurrent users cloning from T0, updating, and merging back.}
  \label{fig:collab-timeline}
\end{figure}

This experiment simulates a collaborative workload when a team of 
data engineers are working on the same dataset.  Four data engineers 
forked their own copies of the lineitem table, T1, T2, T3, and T4 
respectively.  Each engineer updates their own table and in this 
experiment, we assume they cooperate with each other
so that updates of one engineer do not overlap 
with those of another.  This is a commonly used strategy for data engineers
to divide their works into non overlapping sub tasks.
After all the updates are applied, each engineer merges their changes 
back to the original lineitem table.  Figure \ref{fig:collab-timeline} 
shows the timeline of the experiment.  

Table \ref{tab:expdiff4} shows the average time of the four engineers 
to run diff before merging their changes back to the original 
lineitem table.  Table \ref{tab:expmerge4} shows the average 
time to perform the merge.

\begin{table}[h]
  \centering
  \caption{Diff, BuiltIn vs.\ SQL Time(s)}
  \label{tab:expdiff4}
  \begin{tabular}{lcccc}
    \toprule
    \textbf{Operation} & \textbf{C1} & \textbf{C2} & \textbf{C3} & \textbf{C4} \\
    \midrule
    BuiltIn Diff, PK  & 0.08 & 0.28 & 1.36 & 3.04 \\
    BuiltIn Diff, NoPK  & 0.65 & 28.34 & 8.41 & 58.72 \\
    SQL Diff, PK  & 438.41 & 460.33 & 483.35 & 507.52 \\
    SQL Diff, NoPK  & 433.28 & 448.88 & 443.68 & 444.54 \\
    \bottomrule
  \end{tabular}
\end{table}

\begin{table}[h]
  \centering
  \caption{Merge, BuiltIn vs.\ SQL Time(s)}
  \label{tab:expmerge4}
  \begin{tabular}{lcccc}
    \toprule
    \textbf{Operation} & \textbf{C1} & \textbf{C2} & \textbf{C3} & \textbf{C4} \\
    \midrule
    BuiltIn Merge, PK  & 0.26 & 1.69 & 7.98 & 22.81 \\
    BuiltIn Merge, NoPK  & 0.56 & 28.29 & 11.37 & 66.59 \\
    SQL Merge, PK  & 421.87 & 442.96 & 465.11 & 488.36 \\
    SQL Merge, NoPK  & 421.87 & 437.06 & 431.99 & 432.84 \\
    \bottomrule
  \end{tabular}
\end{table}

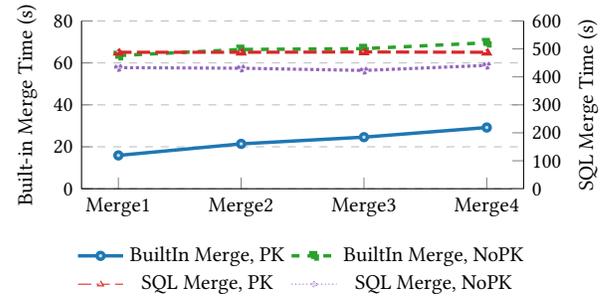
\begin{figure}[h]
  \centering
  \definecolor{BuiltInPKColor}{HTML}{1F77B4}
  \definecolor{BuiltInNoPKColor}{HTML}{2CA02C}
  \definecolor{SQLPKColor}{HTML}{D62728}
  \definecolor{SQLNoPKColor}{HTML}{9467BD}
  \pgfplotsset{
    tablefiveplot/.style={
      width=0.88\linewidth,
      height=0.45\linewidth,
      symbolic x coords={Merge1,Merge2,Merge3,Merge4},
      xtick=data,
      xminorticks=false,
      tick label style={font=\small},
      label style={font=\small},
    }
  }
  \begin{tikzpicture}
    \begin{axis}[
      tablefiveplot,
      axis y line*=left,
      axis x line*=bottom,
      ymin=0,
      ymax=80,
      ylabel={Built-in Merge Time (s)},
      ymajorgrids,
      grid style={gray!50,dashed},
      legend columns=2,
      legend style={
        font=\small,
        draw=none,
        fill=none,
        at={(0.5,-0.28)},
        anchor=north,
      },
    ]
      \addplot+[
        color=BuiltInPKColor,
        mark=*,
        mark options={scale=0.65, fill=BuiltInPKColor!25},
        line width=1.3pt,
      ]
        coordinates {(Merge1,15.9) (Merge2,21.4) (Merge3,24.6) (Merge4,29.2)};
      \addlegendentry{BuiltIn Merge, PK}
      \addplot+[
        color=BuiltInNoPKColor,
        mark=square*,
        mark options={scale=0.6},
        dashed,
        line width=1.3pt,
      ]
        coordinates {(Merge1,63.5) (Merge2,66.4) (Merge3,66.8) (Merge4,69.6)};
      \addlegendentry{BuiltIn Merge, NoPK}
      \addlegendimage{line legend, color=SQLPKColor, mark=triangle*, mark options={scale=0.65, fill=SQLPKColor!25}, dash pattern=on 5pt off 2pt}
      \addlegendentry{SQL Merge, PK}
      \addlegendimage{line legend, color=SQLNoPKColor, mark=*, mark options={scale=0.5, fill=SQLNoPKColor!25}, densely dotted}
      \addlegendentry{SQL Merge, NoPK}
    \end{axis}
    \begin{axis}[
      tablefiveplot,
      axis y line*=right,
      axis x line=none,
      ymin=0,
      ymax=600,
      ytick distance=100,
      ylabel={SQL Merge Time (s)},
    ]
      \addplot+[
        color=SQLPKColor,
        mark=triangle*,
        mark options={scale=0.65, fill=SQLPKColor!25},
        dash pattern=on 5pt off 2pt,
        line width=1.2pt,
      ]
        coordinates {(Merge1,488) (Merge2,488) (Merge3,489) (Merge4,488)};
      \addplot+[
        color=SQLNoPKColor,
        mark=*,
        mark options={scale=0.5, fill=SQLNoPKColor!25},
        densely dotted,
        line width=1.2pt,
      ]
        coordinates {(Merge1,433) (Merge2,431) (Merge3,423) (Merge4,441)};
    \end{axis}
  \end{tikzpicture}
  \caption{Merges (1M updated rows without conflicts).}
  \label{fig:1M_merge_no_conflict}
\end{figure}

The first merge from \texttt{T1} to \texttt{lineitem} table, $\Delta_{lineitem}$ is empty. 
Later merges have more and more rows in $\Delta_{lineitem}$.  Built-in merges take more
time because they need to process more data in later merges.  Figure \ref{fig:1M_merge_no_conflict}
shows the time for merges in the C4 test case (1M updated rows).  The merge times using SQL is 
dominated by scaning the two large tables and there is no observable difference between
merges.  In all four merge operations, builtin merge with primary keys are significantly 
faster than other merge operations.

\subsection{Experiment 4: Collaborative Workload With Conflicts}
We perform the same experiments as in Experiment \ref{sec:expcollab} 
but this time, we let updates of two engineers have 10\% 
overlap on the primary key with each other.  In this experiment, 
the later merges have real conflicts and we let merges to 
resolve the conflicts automatically using the \texttt{ACCEPT} mode.
Table \ref{tab:expdiff42} and Table \ref{tab:expmerge42} shows the average
time for diffs and merges.
\begin{table}[h]
  \centering
  \caption{Diff, BuiltIn vs.\ SQL Time(s)}
  \label{tab:expdiff42}
  \begin{tabular}{lcccc}
    \toprule
    \textbf{Operation} & \textbf{C1} & \textbf{C2} & \textbf{C3} & \textbf{C4} \\
    \midrule
    BuiltIn Diff, PK  & 0.18 & 0.28 & 1.52 & 3.12 \\
    BuiltIn Diff, NoPK  & 0.59 & 7.95 & 8.69 & 61.07 \\
    SQL Diff, PK  & 456.37 & 479.19 & 503.15 & 528.30 \\
    SQL Diff, NoPK  & 445.42 & 448.96 & 456.04 & 454.27 \\
    \bottomrule
  \end{tabular}
\end{table}

\begin{table}[h]
  \centering
  \caption{Merge, BuiltIn vs.\ SQL Time(s)}
  \label{tab:expmerge42}
  \begin{tabular}{lcccc}
    \toprule
    \textbf{Operation} & \textbf{C1} & \textbf{C2} & \textbf{C3} & \textbf{C4} \\
    \midrule
    BuiltIn Merge, PK  & 0.25 & 1.27 & 8.36 & 23.41 \\
    BuiltIn Merge, NoPK  & 0.58 & 8.05 & 11.04 & 66.74 \\
    SQL Merge, PK  & 473.96 & 497.65 & 522.54 & 548.66 \\
    SQL Merge, NoPK  & 448.30 & 451.87 & 459.00 & 457.21 \\
    \bottomrule
  \end{tabular}
\end{table}

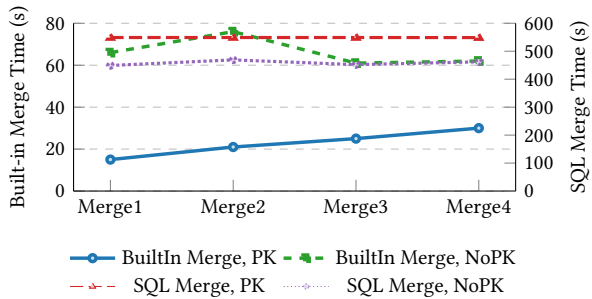
\begin{figure}[h]
  \centering
  \definecolor{BuiltInPKColor}{HTML}{1F77B4}
  \definecolor{BuiltInNoPKColor}{HTML}{2CA02C}
  \definecolor{SQLPKColor}{HTML}{D62728}
  \definecolor{SQLNoPKColor}{HTML}{9467BD}
  \pgfplotsset{
    tablefiveplot/.style={
      width=0.88\linewidth,
      height=0.45\linewidth,
      symbolic x coords={Merge1,Merge2,Merge3,Merge4},
      xtick=data,
      xminorticks=false,
      tick label style={font=\small},
      label style={font=\small},
    }
  }
  \begin{tikzpicture}
    \begin{axis}[
      tablefiveplot,
      axis y line*=left,
      axis x line*=bottom,
      ymin=0,
      ymax=80,
      ylabel={Built-in Merge Time (s)},
      ymajorgrids,
      grid style={gray!50,dashed},
      legend columns=2,
      legend style={
        font=\small,
        draw=none,
        fill=none,
        at={(0.5,-0.28)},
        anchor=north,
      },
    ]
      \addplot+[
        color=BuiltInPKColor,
        mark=*,
        mark options={scale=0.65, fill=BuiltInPKColor!25},
        line width=1.3pt,
      ]
        coordinates {(Merge1,15) (Merge2,21) (Merge3,25) (Merge4,30)};
      \addlegendentry{BuiltIn Merge, PK}
      \addplot+[
        color=BuiltInNoPKColor,
        mark=square*,
        mark options={scale=0.6},
        dashed,
        line width=1.3pt,
      ]
        coordinates {(Merge1,66) (Merge2,76) (Merge3,61) (Merge4,62)};
      \addlegendentry{BuiltIn Merge, NoPK}
      \addlegendimage{line legend, color=SQLPKColor, mark=triangle*, mark options={scale=0.65, fill=SQLPKColor!25}, dash pattern=on 5pt off 2pt}
      \addlegendentry{SQL Merge, PK}
      \addlegendimage{line legend, color=SQLNoPKColor, mark=*, mark options={scale=0.5, fill=SQLNoPKColor!25}, densely dotted}
      \addlegendentry{SQL Merge, NoPK}
    \end{axis}
    \begin{axis}[
      tablefiveplot,
      axis y line*=right,
      axis x line=none,
      ymin=0,
      ymax=600,
      ytick distance=100,
      ylabel={SQL Merge Time (s)},
    ]
      \addplot+[
        color=SQLPKColor,
        mark=triangle*,
        mark options={scale=0.65, fill=SQLPKColor!25},
        dash pattern=on 5pt off 2pt,
        line width=1.2pt,
      ]
        coordinates {(Merge1,549) (Merge2,549) (Merge3,549) (Merge4,548.66)};
      \addplot+[
        color=SQLNoPKColor,
        mark=*,
        mark options={scale=0.5, fill=SQLNoPKColor!25},
        densely dotted,
        line width=1.2pt,
      ]
        coordinates {(Merge1,449) (Merge2,469) (Merge3,452) (Merge4,462)};
    \end{axis}
  \end{tikzpicture}
  \caption{Merges (1M updated rows with conflicts).}
  \label{fig:1M_merge_has_conflict}
\end{figure}

Built-in diff and merge operations performed equally well with the 
no-conflict experiments.  $\Delta$s scanned and aggregated roughly the 
same amount of rows and conflict resolution do roughly the same amount of
work regardless of whether the row had a true conflict.  In fact, in the
case with primary key, there are slightly smaller number of rows after 
diff aggregation than that in the case without conflicts as in Section \ref{sec:expcollab},
 because updates on the same primary key are collapsed.  However, the difference of the
 number of rows in the diff aggregation result is not significant in this experiment setting.
Merging using SQL on the other hand, indeed need to perform more works. 
The delete query using SQL now will need to delete overlapping rows from 
earlier merges.  Figure \ref{fig:1M_merge_has_conflict}
shows the time for merges in the C4 test case (1M updated rows) with conflicts. 

We have experimented diff and merge operations with and without primary key,
with and without conflicts.  Built-in diff and merges performed well across 
all test cases with different change sizes, many times faster than using SQL.  
For a big change set with even a million updated rows, built-in diff and merge 
took only seconds to complete.  This performance advantage is critical for 
collaboration between team members.

\section{Conclusion and Future Work}
The AI revolution is driving new requirements and challenges 
to data engineering.  Software engineers have been using version control 
systems to manage their code and have developed tools and best practices 
for collaboration.  We have no doubt that the same best practices will be
needed and adopted by data engineers to manage their data.  
Data engineers will demand powerful tools for data version control 
so that they can follow these best practices in their daily work.

This paper presents a version control system for data built on MatrixOne's 
snapshot mechanism, which leverages immutable storage and MVCC to enable 
git-like operations on database tables. Our system supports the complete 
set of version control operations—clone, snapshot (tag), diff, merge, and 
revert—on terabyte-scale datasets. Experimental evaluation in Section \ref{sec:eval} 
demonstrates that our implementation of clone, diff, merge are much more 
efficient than equivalent SQL-based implementations.  Data engineers can 
perform these operations on large amount of data   
in seconds, enabling real-time collaboration on large datasets.

Thus MatrixOne enables data engineers to adopt established software engineering 
workflows. Engineers can fork tables to create isolated branches for 
experimentation, use \texttt{SNAPSHOT DIFF} to review changes between versions, 
and merge verified modifications back to the main branch. Multiple team members 
can work concurrently on the same dataset: they can either collaborate directly 
on shared tables with database transactions ensuring consistency, or work in 
isolated forks and merge changes through pull requests. All changes can be 
deployed to production in atomic transactions, ensuring data integrity and 
avoiding service disruptions. This workflow supports continuous integration 
and deployment (CI/CD) pipelines for data, where each modification can be 
traced, validated, and rolled back if needed.
 
Section \ref{sec:discussion} discussed some interesting issues 
and possible improvements.  Better or smarter conflict 
resolution strategies are one of the important areas to work on.
We will continue to work with customers with real-world use 
cases to further improve our version control system.  


\renewcommand\bibname{References}
\bibliographystyle{abbrv}

\begin{thebibliography}{10}

\bibitem{ColumnStore}
Mike Stonebraker, Daniel J. Abadi, Adam Batkin, et al. C-Store: A Column-oriented DBMS.
\newblock {Proceedings of the 31st VLDB Conference, 2005}

\bibitem{CVS}
Concurrent Versions System.
\newblock {https://www.nongnu.org/cvs/}

\bibitem{Git}
Git.
\newblock {https://git-scm.com}

\bibitem{GitLFS}
Git Large File Storage (git-lfs).
\newblock {https://github.com/git-lfs/git-lfs}

\bibitem{GitHub}
GitHub.com
\newblock {https://github.com/}

\bibitem{LSFS}
Mendel Rosenblum, John K. Ousterhout.
The design and implementation of a log-structured file system.
\newblock {ACM Transactions on Computer Systems, Vol 10, Issue 1, 1992}

\bibitem{LSMTree}
Patrick O'Neil, Edward Cheng, Dieter Gawlick, Elizabeth O'Neil.
The Log-Structured Merge-Tree (LSM-Tree)
\newblock {Acta Informatica, Vol 33, Issue 4, 1996}

\bibitem{MatrixOne}
MatrixOne
\newblock {https://www.github.com/matrixorigin/matrixone}

\bibitem{PostgreSQL}
PostgreSQL
\newblock {https://www.postgresql.org/}

\bibitem{Raft}
Diego Ongaro and John Ousterhout.  In Search of an Understandable Consensus Algorithm (Extended Version).
\newblock {https://raft.github.io/raft.pdf}

\bibitem{RCS}
Walter Tichy. Design, implementation, and evaluation of a Revision Control System. 
\newblock{ICSE '82 Proceedings of the 6th International Conference on Software Engineering: 58–67}

\bibitem{S3}
AWS S3.
\newblock {https://aws.amazon.com/s3/}

\bibitem{SCCS}
Marc J. Rochkind. The Source Code Control System
\newblock {IEEE Transactions on Software Engineering, vol. SE-1, no. 4, pp. 364–370, December 1975}

\bibitem{Snowflake}
Snowflake
\newblock {https://www.snowflake.com/}

\bibitem{Supabase}
Supabase
\newblock {https://supabase.com/docs/guides/database/overview}

\bibitem{TPCH}
Transaction Processing Performance Council.
The TPCH Benchmark.
\newblock {https://www.tpc.org/tpch/}

\bibitem{VMDK}
VMware Virtual Disk.  Virtual Disk Format 5.0.
\newblock {http://www.vmware.com/support/developer/vddk/vmdk\_50\_technote.pdf}

\bibitem{ZFS}
J. Bonwick and B. Moore. ZFS: The Last Word in File Systems.
\newblock {http://opensolaris.org/os/community/zfs/docs/zfs\_last.pdf}

\end{thebibliography}

\end{document}